# Self-Assembly of Pheophorbide-*a* on Poly-L–Lysine

**Olga Ryazanova, Victor Zozulya, Igor Voloshin, and Victor Karachevtsev**

*Department of Molecular Biophysics, B. Verkin Institute for Low Temperature Physics and Engineering, National Academy of Sciences of Ukraine, 47 Nauky ave, 61103 Kharkiv, Ukraine.*

[*] Correspondence should be addressed to: ryazanova@ilt.kharkov.ua;

Binding of Pheophorbide-*a* of to poly-L-lysine (PLL) have been studied in neutral aqueous buffered solutions of low and near-physiological ionic strengths with low ethanol content (2.4-5.9 %) a wide range of molar polymer-to-dye ratios (*P/D*) using absorption and polarized fluorescence spectroscopy, fluorimetric titration, fluorescence melting. The binding has highly cooperative character (cooperativity parameter q ≈ 1,000) and results in 40-fold quenching of the dye fluorescence as well as in increase in the fluorescence polarization degree up to 0.16, that evidences formation of π-stacked dye aggregates on the polypeptide exterior. The spectroscopic properties of these aggregates were established. The cooperative biding constants were estimated by Schwarz's method.

# INTRODUCTION

The Pheophorbide-*a* (Pheo-*a*) is natural anionic porphyrin characterized by the chlorine-type visible absorption spectrum with high extinction coefficient in the red region [1] where the transparency of tissues to light increases considerably. This dye is known to be selectively accumulated in the tumor cells being incorporated into their nuclei, mitochondria and lysosomes [2]. It posses by high photosensitizing activity *in vitro* and *in vivo*, that determines its widespread using as a photosensitizer for photodynamical therapy of tumors and psoriasis [3-5]. Also, Pheo-*a* is known to be applied for molecular recognition of G-quadruplexes [6]. The anticancer activity of Pheo-*a* and its possible applications for the treatment of human pathologies were analyzed and summarized in [7,8]. Photophysical properties of Pheo-*a* were reported in [9,10]. It should be noted that the photodynamic activity of Pheo-*a* is mostly determined by monomeric dye molecules, whereas the dimerization [11] or aggregation of the dye in aqueous solutions substantially reduces it. Therefore the peculiarities of the structure and physical properties of complexes formed by this dye with different biological macromolecules found in the blood and body tissue cells are of great interest being subject of intensive study for last many years. The most important of such macromolecules are nucleic acids and proteins. The building blocks of human proteins are 20 amino acids including lysine which is one of two positively charged aminoacids containing additional positively charged aminogroup in its structure which in turn are the binding sites of anionic drug molecules. Lysine promotes the formation of the secondary structure of proteins (the formation of lysine cross-links in collagen, elastin, fibrin and other proteins) through the formation of intra- and intermolecular cross-linking, binding peptide molecules [12]. Lysine is a chiral molecule. In proteins and peptides synthesized in ribosomes, only the L-stereoisomer is found which can form the polypeptides.

Poly-L-lysine (PLL) is a water-soluble linear synthetic homopolypeptide possessing by antimicrobial action [13] which is known to be widely used for cell delivery of nucleic acids and proteins [14-16]. It was one of the first cationic peptides used to mediate gene delivery [17] and as nonviral polymeric gene carriers for a gene therapy [18]. PLL itself is known can be used as an element of biosensors [19–21], whereas its dendrimers were applied for anticancer drug delivery [22] and many others [23–26]. PLL can serve as an excellent system to model properties of the protein structures because depending on the external environments (e.g., solvent and temperature) it can easily adopt three their different most important secondary structures. In aqueous solutions at pH < 10.5 PLL exists as a random coil (the value of pH10.5 corresponds to $pK_a$ of the residual ammonium substituents), at pH >10.5 it takes a conformation of α –helix which transforms into β-sheet when the sample is heated to 50 °C for 10 min [27–29]. High positive charge density of PLL originated from the hydrophilic amino groups on the side chains leads to formation of non-covalent complex between them and negatively charged molecules including DNA [30] and anionic dyes. So, the data on binding of azo-dyes including methyl orange [31–33], ethyl orange [31], and butyl orange [31], cyanine dyes [34], 4'-dimethyl amino azo benzene-4-carboxylic acid [35], tetra-anionic meso-tetrakis(4-sulfonatophenyl)porphine, its zinc(II) derivative [36,37], eosin Y [38] and another ones to PLL are available.

Polarized fluorescent spectroscopy is known as informative and highly sensitive method for investigation of the features of fluorescent dye binding to biological macromolecules. In this work this method along with absorption and melting techniques, was applied to study binding of Pheo-*a* to PLL in buffered aqueous solutions (pH6.9) with low ethanol content (2,4 – 5.9 %) of low (1 mM $Na^+$) and near-physiological (0.15 M $Na^+$) ionic strengths in a wide *P/D* range, and to establish both spectroscopic properties of complexes formed and thermodynamical parameters of binding.

Such a study is very relevant due to the great interest of scientists in porphyrin–polymer hierarchical self-assembly formed by charged porphyrins on oppositely charged polymers where the polymeric matrix acted as a template support for the formation of short- or long-range organized aggregates of chromophores stabilized by porphyrin–porphyrin and porphyrin–polymer weak interactions like van der Waals forces, hydrogen bonding, hydrophobic and electrostatic interactions. The cooperative self-assemblies of porphyrins on biopolymers driven by non-covalent physical interactions (hydrophobic, electrostatic, coordination interaction, hydrogen bonding, and host–guest recognition) turned out to be very promising for building smart soft nanomaterials for optobioelectronics, chiroptical devices, sensors, or materials with extraordinary combination of properties [39,40].

**MATERIALS AND METHODS**
*Chemicals.*

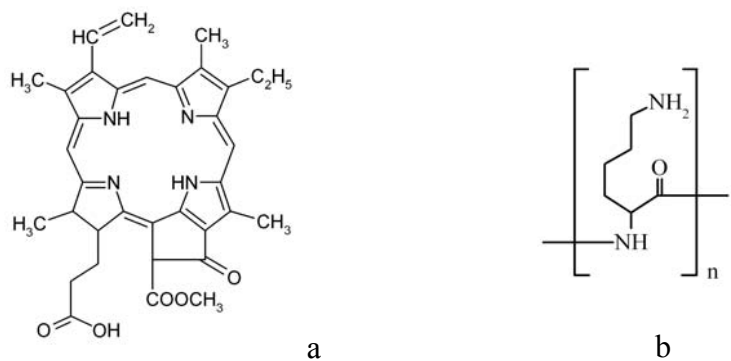

**Figure 1.** Molecular structure of Pheophorbide-*a* and poly-L-lysine (b).

Pheophorbide-*a* (Fig. 1a) from Frontier Scientific Inc. (Logan, Utah, USA) and poly-L-lysine

(Fig. 1b) hydrobromide (Mol. weight is 30,000 - 70,000) from Sigma-Aldrich were used without further purification. The dye concentration was determined spectrophotometrically in ethanol solution using such extinction values: $\varepsilon_{409} = 93000$ M$^{-1}$cm$^{-1}$, $\varepsilon_{667} = 44,500$ M$^{-1}$cm$^{-1}$.

The polypeptide concentrations were determined by gravimetric method.

Since Pheo-*a* has a poor solubility in water, to prepare the samples, the dye was firstly dissolved in ethanol, then water or buffer was added to achieve its desirable concentration (it was in the range $3 \cdot 10^{-6}$ - $1.95 \cdot 10^{-5}$ M). The deionized distilled water and 1 mM sodium cacodylate buffer, pH6.9, prepared from deionized distilled water were used as a solvent.

### *Apparatus and techniques.*

The spectroscopic properties of Pheo-*a* in a free state and bound to PLL have been studied using polarized fluorescence and absorption spectroscopy. The sample solutions were filled in the quartz cells.

Electronic absorption spectra were measured on SPECORD M-40 spectrophotometer (Carl Zeiss, Jena, Germany).

Steady-state fluorescence measurements were carried out on a laboratory spectrofluorimeter based on the DFS-12 monochromator (LOMO, Russia, 350-800 nm range, dispersion 5 Å/mm) by the method of photon counting described earlier in [41]. The fluorescence was excited by polarized beam of He-Ne laser ($\lambda_{exc} = 633$ nm) which radiation was attenuated by neutral density filters. The emission was observed at the 90° angle from the excitation beam. Ahrens prisms were used to polarize linearly the exciting beam as well as to analyze the fluorescence polarization. The spectrofluorimeter was equipped with a quartz depolarizing optical wedge to exclude the monochromator polarization-dependent response. Fluorescence spectra were corrected on the spectral sensitivity of the spectrofluorimeter.

The fluorescence intensity, *I*, and polarization degree, *p*, have been calculated using the formulas [42]:

$$I = I_{II} + 2I_{\perp} \tag{1}$$

$$p = \frac{I_{II} - I_{\perp}}{I_{II} + I_{\perp}} \tag{2}$$

where $I_{II}$ and $I_{\perp}$ are intensities of the emitted light, which are polarized parallel and perpendicular to the polarization direction of exciting light, respectively.

Binding of the Pheo-*a* to PLL was studied using fluorimetric titration in the wide range of molar polymer-to-dye ratios, *P/D*, from 0 to 10.000. Wherein the dependencies of the dye fluorescence intensity and polarization degree on *P/D*, were registered near the maximum of the free dye uncorrected emission band ($\lambda_{obs} = 656$ - 660 nm). The measurements were carried out under the constant dye concentration, for what the dye sample was added with increasing amounts of the concentrated stock solution of PLL containing the same dye content.

For fluorescence melting experiments the cuvette with the sample was placed in a special copper holder of thermal cell which temperature was operated by Peltier element in the range of 20 –99 °C. The cell was inserted into the spectrofluorimeter.

All experiments were carried out at ambient temperature (20 – 22 °C).

**RESULTS AND DISCUSSION**

***Absorption and fluorescence spectra of Pheo-a in a free state and bound to poly-L-lysine***

Absorption spectra of free Pheo-*a* in ethanol and water with 5.34 % of ethanol are presented in Figure 2. The spectrum in ethanol is known to be attributed to the dye monomers. In the visible region it consists of the intense Soret band (or B-band with high exctinction coefficient, $\varepsilon_{409}$ = 93,000 $M^{-1}cm^{-1}$) with maximum at 409 nm, and four Q-bands at 507 ($Q_x(0,1)$), 538($Q_x(0,0)$), 609 ($Q_y(0,1)$) and 667 nm ($Q_y(0,0)$), among which the last one is most intensive ($\varepsilon_{667}$ = 44,500 $M^{-1} cm^{-1}$) that is typical for chlorine derivatives [1,43]. All these bands correspond to $\pi \rightarrow \pi^*$ transitions polarized within the plane of conjugated macrocycle.

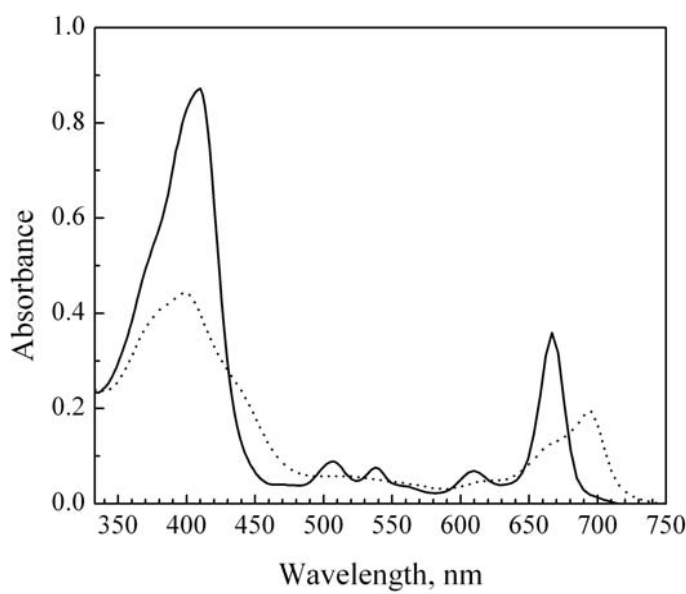

**Figure 2.** Absorption spectra of Pheo-*a* in ethanol (solid line) and in 1 mM aqueous buffered solution with 5.34 % of ethanol (dashed line), $C_{dye}$ = 10 µM, path length 1 cm.

In aqueous solution with low ethanol content Pheo-*a* absorption spectrum undergoes approximately 50 % hypochromism, *ca.* 10 nm blue shift of the Soret band maximum (up to 392 – 397 nm depending on the dye concentration and ethanol content) and appearance of a shoulder in its red side (Figure 2). For longwave absorption band substantial broadening was observed along with large red shift (*ca*. 30 nm) of its maximum (up to 696 – 699 nm depending on the dye concentration and ethanol content). The band becomes superposition of the two unresolved band: maximum of the first of them corresponds to that in ethanol and the second one is a new intense band with maximum at approximately 697 nm. Since it is known that in aqueous solution at the concentration studied hydrophobic Pheo-*a* tends to form dimers [44], absorption spectrum is obviously corresponds to monomer-dimer mixture.

The fluorescence spectrum of Pheo-*a* in the aqueous solution represents a broad asymmetric structureless band ($\lambda_{max}$ = 660 nm) with gentle slope of its longwave part (the shoulder). The fluorescence polarization degree, *p*, measured at the emission band maximum, was found to be around 0.025.

Absorption and fluorescence spectra of Pheo-*a* + PLL complex with *P/D* = 10,000 were recorded. Its visible absorption bands turns out to be blue shifted in comparison with those for the free dye: maxima of the Soret and longwave absorption bands are located at 385-388 and 686 nm correspondingly. The fluorescence spectrum of this complex represents wide structured band with maximum at 693 nm (700 nm in corrected spectrum), which shape evidences that it is the superposition of at least two constituent bands: the first one centered at approximately 657 nm, which was observed in the spectrum of the free dye, and second new more intense band centered at 693 nm (Figure 3a). For the freshly prepared complex the ratio between them is $I_{657\ nm}/I_{693\ nm}$ = 0.73, whereas 1 hours later (the sample was kept in the dark) it reduces up to 0.67. It can be noted that such change is occurred dye to a decrease in the $I_{657\ nm}$, whereas $I_{693\ nm}$ remains almost unchanged. The values of fluorescence polarization degree measured at 657 and 693 nm are 0.160 and 0.129 correspondingly.

The serial mixing (dilution) of the complex (*P/D* = 10,000) with the free dye solution (*P/D* = 0) was performed to obtain set of the samples with *P/D* lying in the range of 10,000–50 (9 complexes). For complexes with *P/D* =7333, 5000, 3500, 2500 and 1800 the equilibrium was achieved only a few hours after mixing (the samples were kept in the dark) and it was accompanied by the change in the shape of the emission band over time. Example of such change is presented in the Figure 3b for complex with *P/D* = 5000. It is seen that more than 1.5 –fold decrease of the fluorescence intensity of the first constituent band ($\lambda_{max}$ = 657 nm) was observed over time whereas that for the second one ($\lambda_{max}$ = 693 nm) it was practically unaltered. For complexes with *P/D* = 500 and 250 the equilibrium was achieved over 10 – 20 minutes. And for complexes with *P/D* = 125 and 50 the equilibrium was achieved promptly. It should be noted that for the samples with *P/D* < 1800 their emission bands have practically the same shapes and have no feature at 693 nm (Figure 4). The integral intensity of the fluorescence bands for the sample with *P/D* = 10,000 is *ca.* 1.44 and 1.57 times higher than those for the samples with *P/D* = 5,000 and 3,500 correspondingly.

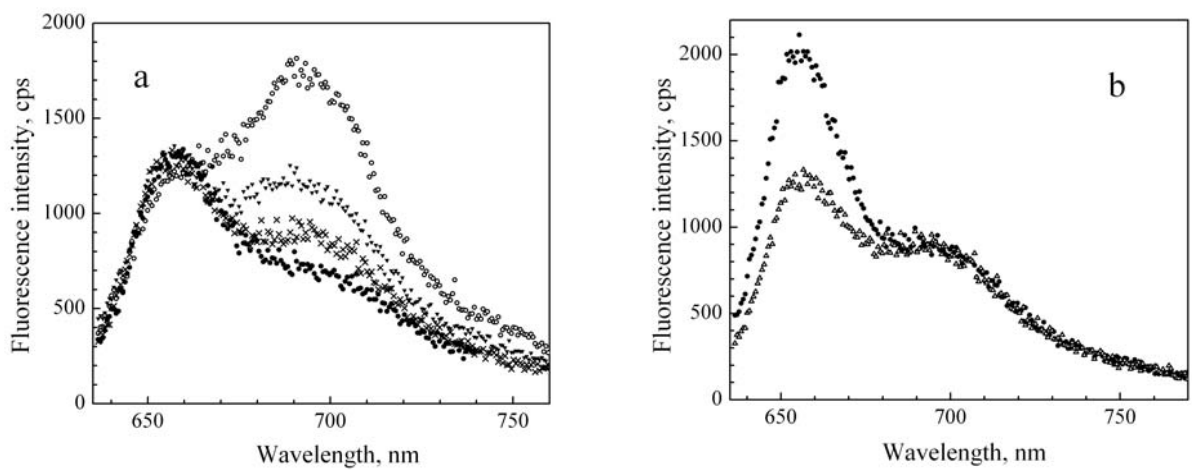

**Figure 3. (a)** Fluorescence spectra of Pheo-*a* complex with poly-L-lysine in 1 mM aqueous buffered solution with 5.9% of ethanol *P/D* = 10000 (o), 7333 (▼), 5000 (×), 3500 (●), $C_{dye}$ = 19.5 μM, $\lambda_{exc}$ = 633 nm.

**(b)** Fluorescence spectra of Pheo-*a* complex with poly-L-lysine (*P/D* = 5000) in 1 mM aqueous buffered solution with 5.9% of ethanol after mixing (●) and a few hours later (o), $C_{dye}$ = 19.5 μM, $\lambda_{exc}$ = 633 nm. The sample was kept in the dark.

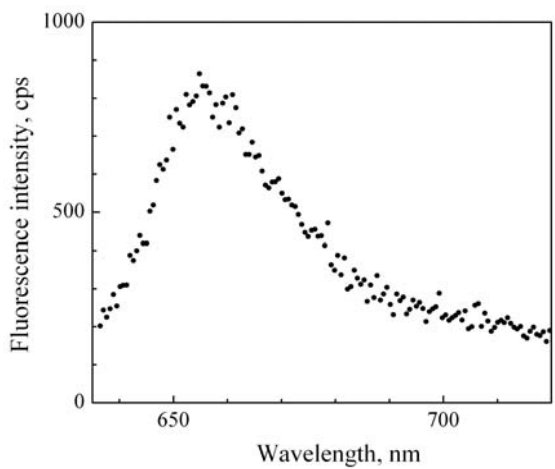

**Figure 4.** Fluorescence spectra of Pheo-*a* complex with poly-L-lysine in 1 mM aqueous buffered solution with 5.9% of ethanol *P/D* = 500 (●), $C_{dye}$ = 19.5 μM, $\lambda_{exc}$ = 633 nm.

Absorption measurements show that the decrease in *P/D* from 10,000 to 125 is accompanied by the moderate hypochromism of the absorption bands as well as by red shift of the longwave absorption band maximum from 686 to 699 nm, whereas position of the Soret band was practically the same. In the *P/D* range from 0 to 50 an increase in the PLL concentration results in the hypochromism and blue shift of the longwave absorption band (Figure 5). As for the Soret band, in the range of 0 to 2 it shows hypochromism and further *P/D* increase induce the rise of its intensity and the narrowing due to disappearance of the longwave shoulder.

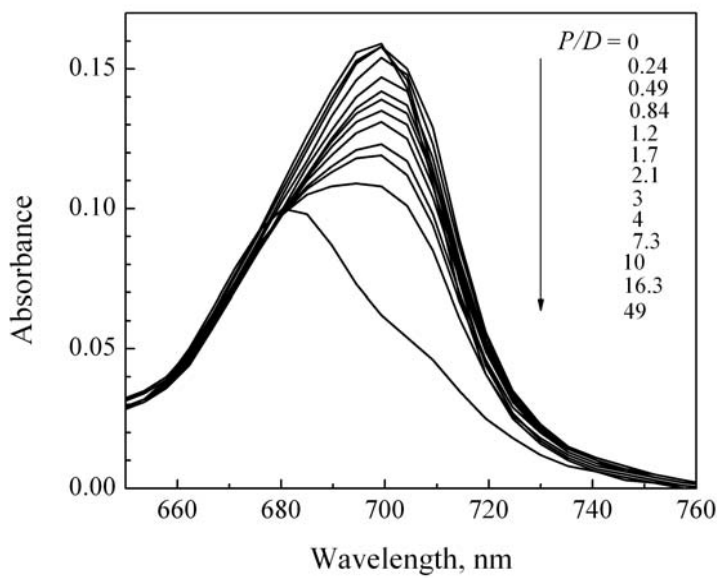

**Figure 5.** Longwave absorption band of Pheo-*a* in a free state and bound to PLL in the range of *P/D* = 0 – 49 measured in aqueous solution with 2.4 % of ethanol, $C_{dye}$ = 3 μM, path length 2 cm.

*Fluorimetric titration study of Pheo-a binding to poly-L-lysine*

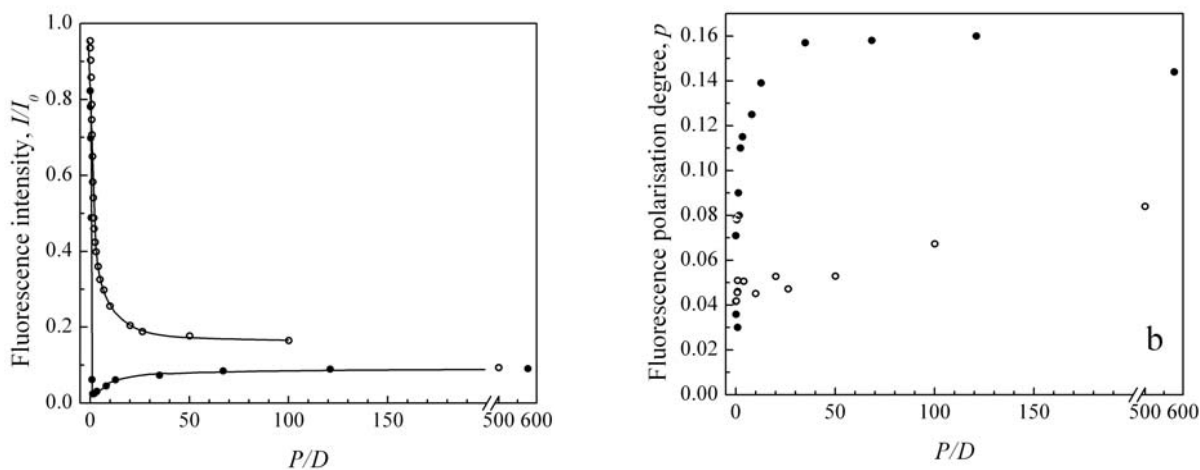

**Figure 6.** Dependence of relative fluorescence intensity (a) and polarisation degree (b) of Pheo-*a* measured in the maximum of the dye emission band on molar *P/D* ratio upon titration with poly-L-lysine in aqueous buffered solution containing 5.6% of ethanol and 1 mM $Na^+$ (●) and 0.15 M $Na^+$ (o), $C_{dye}$ = 1.9·$10^{-5}$ M, $\lambda_{exc}$ = 633 nm, $\lambda_{obs}$ = 655 nm.

The fluorescent titration curves of Pheo-*a* with PLL presented as a dependence of fluorescence intensity (a) and fluorescence polarization degree (b) near the free Pheo-*a* emission maximum on polymer-to-dye ratio are shown in Figure 6. It is seen that for low ionic strength sample containing 1 mM $Na^+$, the strong fluorescence quenching (Figure 6a) and increase in fluorescence polarization degree (Figure 6b) from 0.025 to 0.16 has been observed. From Figure 7 showing initial part of fluorescence titration curve it is seen that the curve consists in three sections. The first one lies in the range of 0 < $P/D$ < 0.5, here the value of the fluorescence intensity is practically unchangeable. The second part observed in the range of 0.5 < $P/D$ < 1 represents linear descending dependence of $I/I_o$ on $P/D$ reaching minimum at *ca.* $P/D$ = 1. And in the range of 0 < $P/D$ < 10 practically constant level of the dye emission was observed. Further increase of the polymer content results in only slight rise of fluorescence intensity at high $P/D$ values (up to 9 % from initial at $P/D$ = 600 (Figure 6a) and to 12 % at $P/D$ = 8000).

It is known that in aqueous solutions at given concentrations the free Pheo-*a* exists mainly in the form of dimers [44] which turn out to be quite stable and do not immediately disintegrate upon the polymer addition, as evidenced by first part of the titration curve. The second linear descending part observed in the range of 0.5 < $P/D$ < 1 is typical for the cooperative binding of both anionic dyes to linear polycations and for cationic dyes to linear polyanions due to their Coulomb interaction [45,46]. So it can be attributed to Pheo-*a* dimer disintegration with consequent electrostatic binding to PLL exterior accompanied by π-stacking of the dye chromophores and formation of weakly- or non-fluorescent aggregates. In other words, Pheo-*a* self-assembly on PLL was occurred. Being extrapolated to the ordinate axis, at $P/D$ = 0 this dependence gives the value of relative fluorescence intensity of Pheo-*a* monomers. At $P/D \approx 1$ the binding saturation and existence of extended porphyrin aggregates was observed, the minimal level of residual fluorescence was about 2.4 % of that for the free monomeric dye. In theory, due to entropy factor, upon increase of the relative polymer content, these aggregates should disintegrate transforming into the bound dye monomers at high $P/D$ values. However, very low level of the dye fluorescence (9% from initial one) registered at $P/D$ = 600 evidenced that complete dissociation of stacks was not occurred and at high polymer content the dye is apparently bound to PLL as a small aggregates (dimers, trimers, tetramers). The increased *p* value (0.16) can be explained by chromophore immobilization on the polycation template upon the porphyrin binding or by decrease in its fluorescence lifetime.

Comparison of the fluorescence titration curves for Pheo-*a* with PLL at low ionic strength with those for monocationic pheophorbide-*a* derivative, CatPheo-*a*, with inorganic polyphosphate, PPS [47], shows that the residual level of fluorescence upon saturation of the polymer lattice by the dye molecules was the same in both cases, being equal to 2.4 % from initial. $I/I_o$ ratios registered at high $P/D$ are also similar, being equal to 9 % and 10 % for Pheo-*a* + PLL and CatPheo-*a* + PPS correspondingly. At the same time the dependence of *p* on $P/D$ for these two systems are different. For Pheo-*a* + PLL the monotonic rise of *p* up to 0.16 is observed upon $P/D$ increase from 0 to 35, than it reach steady level, slightly decreasing to 0.14 at $P/D$ = 600. Whereas for CatPheo-*a* + PPS fluorescence polarization rises sharply up to 0.12 in the range of 0 < P/D < 3, then it gradually decreases to 0.05 at high $P/D$.

It is known that in water at neutral pH7 PLL has a disordered "random coil" structure. Electrostatic binding of anionic Pheo-*a* to cationic poly-L-lysine along with hydrophobic interaction results in self-stacking of the dye chromophores on the polypeptide exterior and formation of the extended aggregates. This is cooperative process that is confirmed by the linear shape of initial part of fluorescent titration curve presented as $I/I_0$ vs $P/D$ dependence [48] (Figure 7). It is supposed that during this process Pheo-*a* induces changes in the PLL conformation from disordered coil to partially ordered linear or helical structure because of the polymer adjustment to pheophorbide stacks.

Increase in the solution ionic strength up to the physiologic value, 0.15 M NaCl, attenuates the electrostatic binding of Pheo-*a* to PLL due to the competitive binding of counterions, that results in another character of fluorescence titration curve (see Figure 6a) where the quenching of Pheo-*a* fluorecence intensity only up to 16 % from initial one is observed. The value of fluorescence polarization degree increases up to $p = 0.08$, which is substantially lower than that in the solution of low ionic strength.

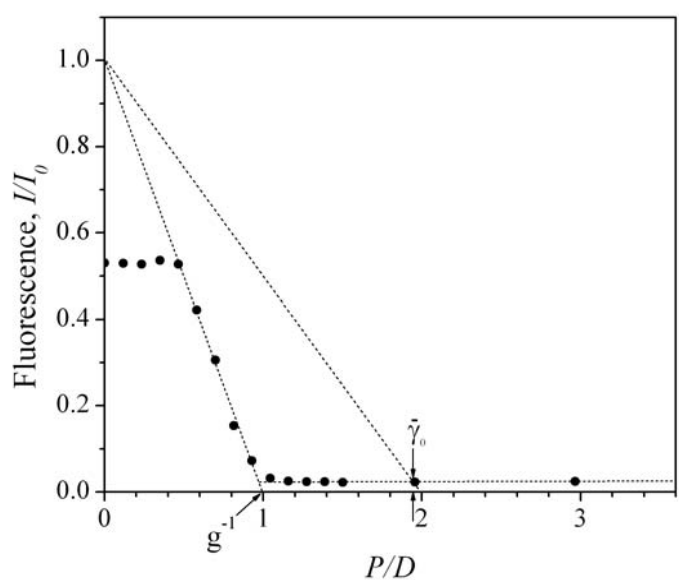

**Figure 7.** Initial part of fluorescent titration curves presented as dependence of relative Pheo-*a* fluorescence intensity on $P/D$ upon titration with poly-L-lysine in 1 mM cacodylate buffer with 6% of ethanol, $C_{dye}$ = 19.5 µM (●), $\lambda_{exc}$ = 633 nm, $\lambda_{obs}$ = 655 nm. The dashed lines are used for determination of parameters $g$ and $K$ (see text)..

*Fluorescence melting study of Pheo-a binding to poly-L-lysine*

To found out whether the complex formed by Pheo-*a* with PLL is stable, melting of the sample ($P/D = 500$) with fluorescence registration was performed. Figure 8 shows the heating and cooling curves registered in the temperature range from 20 to 99 °C. The experiment evidences the high thermal stability of these complexes, which dissociation has been begun only at T = 50 °C, and it has not been completed up to 100 °C. The process is reversible, however a substantial hysteresis was observed.

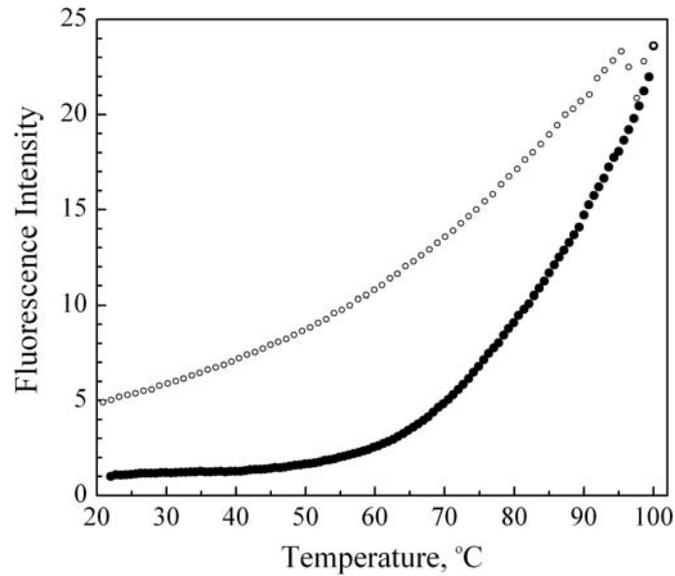

**Figure 8.** Temperature dependence of the Pheo-*a* fluorescence intensity upon dissociation (●) and association (o) of its complex with poly-L-lysine (*P/D* = 500) in aqueous buffered solution containing 1 mM $Na^+$, $C_{dye}$ = 1.95·$10^{-5}$ M, $\lambda_{exc}$ = 633 nm, $\lambda_{obs}$ = 657 nm.

*Parameters of the cooperative binding of Pheo-a to poly-L-lysine*

The initial part of fluorescent titration curves (Figure 7) in the range of 0.5 < *P/D* < 1 shows sharp descending linear dependence of *I/I₀* on *P/D* evidencing that electrostatic binding of monoanionic Pheo-*a* to polycationic PLL is highly cooperative process. Very low level of residual fluorescence allows us to conclude about self-stacking of the dye chromophores on the polypeptide exterior. The curves were used to estimate the thermodynamical parameters of cooperative binding of Pheo-*a* to PLL by Schwarz's method [48]. If to assume that all dye is bound in the self-associated form and fluorescence of the aggregates is negligible, the dependence of *I/I₀* on *P/D* can be considered as dependence of unbound dye fractions, $\gamma_0$, i.e. $\gamma_0$ = *I/I₀*. The straight-line section of this curve corresponds to stoichiometric binding conditions. Being extrapolated to *P/D* axis it gives a binding site value as $\frac{1}{g} \approx 1$ monomeric units of poly-L-lysine per one Pheo-*a* molecule, i.e. number of binding sites per pLL monomer unit is g ≈ 1 (Figure 7).

The cooperative binding constant was estimated according to the equation

$$K = \frac{1}{\bar{\gamma}_0 \cdot C_T} \qquad (2)$$

where $C_T$ is the total dye concentration, and $\bar{\gamma}_0$ is the fraction of free ligand provided the polymer lattice is half-occupied. The $\bar{\gamma}_0$ value was determined as an ordinate of a dashed line interception with the fluorescent titration curve in Figure 7. For solutions of low ionic strength, the estimated value of cooperative binding constants, $K$, obtained in this way was found to be *ca.* $2.1 \cdot 10^6$ M$^{-1}$. Since actually the fluorescence of aggregates was not quenched completely, it makes a some contribution to $I/I_0$ that results in conservative value of $\bar{\gamma}_0$. I.e. the above value of $K$ is the minimal possible value of cooperative binding constant.

The binding cooperativity parameter $q$ was estimated from the relation:

$$K = qK^+ \qquad (3)$$

where $K^+$ is a binding constant of isolated (non-aggregated) ligand molecule. Taking into account the value of $K^+ = 1800$ M$^{-1}$ [49], we have obtained the binding cooperativity parameter, $q \approx 1000$.

The thermodynamical binding parameters obtained for Pheo-*a* +PLL are practically the same as those obtained earlier for complex of CatPheo-*a* with PPS [47], where g ≈ 1, $K = 1.9 \cdot 10^6$ M$^{-1}$, and $q \approx 1000$.

For the sample of near-physiological ionic strength, 0.15 M Na$^+$, the equilibrium binding constant was estimated as 1.4–1.8· 10$^5$ M$^{-1}$, this is one order of magnitude lower than that of low ionic strength, 1 mM Na$^+$, due to competitive binding of counterions to PLL.

**CONCLUSIONS**

Anionic Pheophorbide-*a* exhibits a strong binding affinity to cationic poly-*L*-lysine with high cooperativity of binding process. At low polypeptide content (*P/D* < 10) the dye molecules self-assemble into continuous ordered aggregates on the polymer exterior with π-stacking between the adjacent chromophores that is confirmed by the strong quenching of their fluorescence and increase in the fluorescence polarization degree up to 0.16. The residual emission intensity at *P/D* ≈ 1 is only about 2.4 % from that for the free dye. Self-aggregation of Pheo-*a* is also characterized by hypochromism and blue shift of the visible absorption bands. In the solution of low ionic strength the cooperativity parameter and the cooperative binding constant by Schwarz's method were $q \approx 1000$ and $K \approx 2.1 \cdot 10^6$ M$^{-1}$ correspondingly. Number of binding sites per pLL monomer unit is estimated as g ≈ 1. It is supposed that Pheo-*a* induces changes in the PLL conformation from disordered coil to partially ordered linear or helical structure because of the polymer adjustment to pheophorbide stacks.

At the high pLL content (*P/D* > 100) Pheo-*a* are presumably bound to the poly-*L*-lysine chain as a small aggregates (dimers, trimers, tetramers).

Increase in the solution ionic strength up to the near physiologic value, 0.15 M NaCl, attenuates the electrostatic binding of Pheo-*a* to PLL due to the competitive binding of counterions, that results in the quenching of Pheo-*a* fluorecence intensity only up to 16 % from initial one and rise of the fluorescence polarization degree up to $p$ = 0.08. The equilibrium binding constant estimated as 1.4–1.8· $10^5$ $M^{-1}$, is also one order magnitude lower than that in the case of low ionic strength.

Using the example of poly-L-lysine, we have shown that pheophorbide being introduced into the cell actively binds to oppositely charged proteins. The strong tendency of Pheo-*a* to self-association upon their interaction can decrease photodynamic activity of the dye since dimers or aggregates of this porphyrin exhibit reduced production of the singlet oxygen.

**DECLARATION OF INTEREST STATEMENT**

The authors declare that our manuscript complies with the all Ethical Rules applicable for this journal and that there are no conflicts of interests.